\documentclass[onecolumn,numbook]{svjour2}{}
\usepackage[pdftex]{graphicx}
\usepackage{graphics}
\usepackage{amsmath,amsfonts,amssymb}
\usepackage{longtable}
\usepackage[pdftex]{hyperref}
\journalname{Foundations of Physics} 
\smartqed
\usepackage{mathptmx}      

\begin{document}

\title{A physicist's view of the universe: a philosophical approach}

\author{Israel P\'erez}

\institute{Israel Omar P\'erez-L\'opez \at
              Department of Applied Physics, CINVESTAV Unidad M\'erida \\
              Tel.: +52-(999) 942 94 00 Ext. 2265 \\
              Fax: +52-(999) 981 29 17 \\
              \email{cooguion@yahoo.com, iperez@mda.cinvestav.mx}             \\
            \emph{Present address: Km 6 Antigua Carretera a Progreso
A. P. 73, Cordemex, C.P. 97310, M\'erida, Yucat\'an, M\'exico}}

\date{Received: date / Accepted: date}

\maketitle

\begin{abstract}
Without a doubt many problems in physics arise as a consequence of our philosophical conception of the world. In this contribution however we endeavor to alleviate this scenario by putting forward a philosophical approach under which some of the most fundamental problems in modern physics might turn out to be fictitious. To accomplish such task we propound that everything that exists must be made up of \emph{matter} which not only makes up space and the universe but also is in constant change. For such reason the existence of total emptiness and material discontinuity are rejected. Here physical fields are assumed as a particular state of matter. And time is understood as the result of the intrinsic dynamics of the universe. Furthermore, the infiniteness of the universe is also discussed and its implications are briefly mentioned, e.g., the laws of conservation. Finally, the regularity of the physical laws is questioned. In summary four great problems (from the perspective of physics) are suggested to be deeply studied: (1) What is matter?, (2) Why does the universe change? (3) is the universe infinite in extension? and (4) are there really regular (invariant) laws of physics?

\keywords{Matter \and Continuum \and Time  \and Infinite Universe \and Regular Laws \and Total Knowledge}
 \PACS{ 01.65.+g \and 01.70.+w \and 98.80.-k   \and 98.80.Bp }
\end{abstract}

\section{Introduction}

As it is well known, in order for a physical theory to be successful certain requirements must be fulfilled \cite{popper1}. (a) A theory must not only be as simple as possible (Occam's razor) but also (b) must be founded on an axiomatic formulation. (c) Mathematical elegance is another desired quality that might stem from the previous items; by this we also mean that theories must be mathematically consistent. (d) However, according to Popper \cite{popper1}, a hypothesis can be considered as scientific only if it is falsifiable. (e) But above all, the new theory must not only explain the phenomena that the old theory explains but further make new predictions. Such predictions must be, in the particular case of physics, experimentally testable and when this process is carried out we must avoid, again according to Popper, evading and deceiving the falsifiable criteria. In practice, such criteria are not only hard to carry out but also difficult to identify because most of the times the objects of study in physics can only be scrutinized by indirect methods of observation (measurements), that is, by the analysis of the correlations among the different physical quantities (variables, observables, parameters). In essence, the theory of relativity and quantum mechanics gained their status complying some of the above virtues. Thus, for if unified theories, such as M-theory, are to be successful they must not only comprehend or encompass the latter theories but also make new predictions. (f) Another important factor that is commonly ignored by physicists and eclipsed by the above items is the coherence in the physical interpretation, by this I mean, the epistemological coherence that appears when we try to decode the mathematical language and put it into ordinary and intuitive language; Max Tegmark refers to this as \emph{baggage} \cite{tegmark5}, but I would rather punctually say \emph{natural philosophy}. 

Very recently, as a consequence of the popular unified theories \cite{smolin1,penrose,greene1}, the latter factor has regained major importance. To make my view clearer, let us consider the case of M-theory which requires, to be mathematically consistent, the existence of ten dimensions. Being honest we should admit that it is very difficult to reconcile ourselves with the idea that there are more than three spatial dimensions. So, one may ask: what physical evidences support their existence? What powerful epistemological \footnote{Obviously, many metaphysical arguments can be raised to justify $n$-dimensional universes.} reasons do we have to believe in extra dimensions? Yet I am far from agreeing with all those physicists who favor the feeble argument of the cable that is seen from a very far distance and appears to be of one dimension, the closer we look at the cable the more dimensions we observe, they say. For if the incorporation of dimensions is just a mere mathematical artifice that only frees us from major complications of the same nature, it seems to me more plausible that we reanalyze our intuitive vision of the universe and put the feet on the ground before accepting such proposals that, in spite of that such theory widely fulfills the expectations of items (b), (c) and (e), they leave much to be desired from items (a), (d) and (f). 

Besides, one should recall that physics is not pure mathematics, physics makes the connection between the abstract universe and the real one (or the measurable universe if you wish), physicists grant tangible sense to mathematics and, at the same time, describe the real universe by means of sets of mathematical symbols, that is, by physical laws. Thus, if our mathematical theories are to describe the real universe we should acknowledge that mathematics is an experimental science, otherwise the practice of math is mere metaphysics without pragmatic usage for real life. 

In order to carry out such a task, we must, it seems to me, renew the epistemology of physics, revive the philosophical spirit and, thus, recover the tradition practiced by the physicists of the previous centuries. The way physics is carried out today is so abstract that the physical sense and the intuitive notions are almost lost. And I think that another way of growing our understanding of the universe cannot only be attained by abstract theories and experimental observations but by philosophical reasoning as well. Hence, if the reader has captured my sketch he will realize that that is the aim of the present contribution. I must make clear that my objective is not to establish precise physical or mathematical definitions of what we shall treat here, but, departing from physical and philosophical principles, to put on the table, based on logic, problems that, under the judgment of the author, are some of the most essential that contemporary physics must profoundly understand if great advances upon the knowledge of the universe are desired. I must warn the reader that the proposal to be developed in the following pages does not stand somewhat allied to the established corpus of physics, but, however, it can be of great aid to get to the bottom of some of the most fundamental puzzles in physics. 

One of the main aims of this work is to expose the intuitive perspective that I envisage of the cosmos based on my own experience in physics and on  ``common sense". Thus, I shall endeavor to show that from some natural assumptions and reasonings valuable physics can be extracted eluding the complications of the mathematical approaches.

During the exposition of this article we shall realize that some problems of modern physics may appear somewhat to be fictitious for they might arise from our philosophical conception of the world. In the following section I shall list some of the most relevant problems in contemporary physics. Later in section \ref{univer} I shall expose my conception of the universe. Whilst in section \ref{sec} we shall devote some lines to discuss which of these problems might be addressed (or tackled) following our philosophical proposal. Finally, some concluding remarks are given in section \ref{conclu}.

\section{The trouble with physics}
\label{trou}

Now let us list some of the problems that contemporary physics regards as the most fundamental \footnote{Certainly, some other great problems have been omitted since they are outside the scope of the present investigation.}. We can summarize them as follows:
\begin{enumerate}
  \item The unification of the general relativity and quantum mechanics.
  \item The foundational problems of quantum theory (relation observer-system, the collapse of the wave-function, wave-particle duality, paradoxes and the underlying reality).
  \item The unification of particles and interactions (forces) as manifestation of a single entity.
  \item What are dark matter and dark energy? Why is the expansion of the universe accelerating? 
   \item Higgs mechanism, does the Higgs particle exist? 
\item Galaxy rotation problem.
   \item Hipparcos, Pioneer and Fly-by anomalies.
  \item Existence of multiple universes.
  \item Existence of extra dimensions.
  \item  The arrow of time (symmetry of the laws of physics before time reversal).
    \item Explanation of why the parameters in the standard model of particles have that value.
   \item Explanation of why the constants in the standard model of cosmology have that value.
 \item Is it possible a theory of everything?
\end{enumerate}

\section{The Universe}
\label{univer}

\begin{enumerate}

 \subsection*{The universe is absolute}
 \item I shall speak in absolute terms because it is absolute that there is something that continuously stimulates our senses. That one that denies it, I think, contradicts himself by doing it, for, at least, he must acknowledge that he is capable of perceiving things regardless of the nature of their origin, that is, whether they are material, spiritual, physical, real, volitive, metaphysical, exist or not... and this is absolute. But here, I shall constrain myself only to talk about something that we ``believe" is physical because exists outside our minds with absolute independence and is made up of certain substance(s) with a complex dynamics to be known; that something is called the \textsc{universe}.
  
  \item Some philosophers \cite{moore1} argue that there is not any proof of the existence of an external world. But it seems to me that we cannot deny that we are capable of perceiving that we are something no matter what we are. Those who assert that we cannot be sure of anything contradict themselves, for, at least, they are sure that one cannot be sure of anything. In this sense, the previous statements lead us again to the notion of an absolute, either one position or the other is adopted, whichever of them represents an absolute position. Whether the universe is external and independent of our existence or not; or whether we are real or not, that is, for our purposes, inconsequential, what matters is that there is absolutely something that we feel, think and sense, and it must be investigated following the laws of the reasoning whichever they are \cite{mittelstaedt}.

\subsection*{Thinking as a result of the evolution of the universe}
  \item From a scientific perspective it has been said that thought is the result of the evolution of the universe and under this premise physical theories are created. What has been said will awake, in some, the idea of the anthropic principle \cite{smolin}; in others it will bring into the mind the object-subject problem from the theory of knowledge \cite{hessen} where the object to be known is the universe; for some others the monism, the dualism, the pluralism, the idealism, etc.; or perhaps, the idea of an intelligent designer; and in some others, the idea of being part of a computer simulation created by a biological primitive race of super-intelligent  beings \cite{schmidhuber,mccabe,bostrom}. It is not significant how we address our problem, a healthy human mind in any of these scenaria will demand questions of similar nature: What is the universe? How does it work? Is it possible to thoroughly know and understand the universe? We all realize that such straightforward questions are the most difficult to answer in physics. Answering these questions implies the finding of the master equation and the theoretical system that convince and satisfy to the laws of reasoning and logic. Such system physicists call it the \emph{theory of everything} (TOE) \cite{tegmark5,tegmark1}, philosophers, however, argue that there are limits to knowledge \cite{moore2}.
 
  \subsection*{The problem of knowledge}
  
  \item And when I say ``what is the universe?" I mean, how we understand everything we perceive and feel, how we interpret those sensations and experiences, how we represent the interaction with the universe in terms of models and theories. Such interpretations are what we call knowledge. Thus, the trouble resides on that we do not know yet the entire universe and that we do not know whether is possible to have access to the totality. What apparently is true is that our knowledge of the universe has grown. But, I ask myself: will it be a maximal knowledge? Or better yet, is it really possible the knowledge? And certainly, according to Popper \cite{popper1}, there exists the problem of knowing more, the central problem of epistemology, the problem of the growing of knowledge. Descartes realized that as his knowledge grew, he became more perfect \cite{descartes2,descartes3,descartes5} but what sense would our existence acquire whenever we reached the total and absolute knowledge of the universe? would we become omniscient and omnipotent gods? Would it be possible to reach such point, the Omega point of Teilhard de Chardin? If not, what is the fundamental mission of a physicist?

\subsection*{Where does the universe come from? What is the universe made up? And what are space and time?}

\item Now, if I am something I must be made up of something that exists in itself, absolute and independent of my mind. Similarly, if the universe is something it must be made up of something, of some substance. Such substance we can call it \emph{matter}\footnote{Another appropriate name can be \emph{energy} but due to the relation $E=mc^2$, matter is not a derived entity, so, for convention, we can choose matter as a primordial entity.} (M) [in contemporary physics fields are regarded as something of different nature than matter, however, here we shall conceive a field as some kind of matter in certain state (please see \cite{perez1,volovik,christov7,christov5,christov1,christov6})]. The study of matter constitutes our first great fundamental problem. One of the properties of matter is \emph{mass}, and mass, as it is well known, is a source of gravitation and, at the same time, is some kind of \emph{energy} \cite{jammer1}. Though there remain to thoroughly understand what these things are and how they fundamentally interact. From here it follows that if space and time \footnote{Here we allude to the Newtonian \cite{newton1} notion of space and time, i.e. for Newton space and time existed by themselves as the container of matter and events, respectively. But space and time were not material entities. The Einsteinian notion is omitted to avoid confusion (please see also the notions according to Immanuel Kant \cite{kant} and Max Jammer \cite{jammer2}).} are physical entities they must be made of M, otherwise their nature might be emergent, associative  and/or relational as \emph{taste} or \emph{smell}. Furthermore, we should admit that, according to experience and common sense, such substance might be in perpetual change and motion, in inexorable mutation. If this is true, it might be then that time is one of its consequences \cite{mitski,zeh,barbour,markopoulou}.

\item For our purposes we can classify matter in, at least, two manifestations as it was conservatively realized in the XIX century. There exists ponderable matter\footnote{Bulk and gross matter are also assumed for this manifestation of matter.} which comprehends all solids, fluids, plasmas and particles (From the perspective of physics this matter corresponds to the set of particles of the standard model); and \emph{imponderable matter} which is that one that James C. Maxwell discussed in his works \cite{max1,max2,max3}. We understand imponderable matter as that primordial matter that constitute a continuum \cite{christov7,christov5,christov1,christov6} and is the progenitor of ponderable matter and, among other things to be investigated, serves to propagate the interactions (force fields). Some call it the aether, spirit, space, others quantum foam, others the metacontinuum, modern physics call it the quantum vacuum (though currently it is not seen as something material), background, the Higgs field, etc. Here the name is irrelevant, what is important of this is the notion that there is a subtle continuous material entity that makes up the universe \cite{volovik}.

  \item Moreover, I support the idea that something cannot be created out of nothingness; understanding nothingness as the absence of any kind of M. From this affirmation it follows that it is useless to inquire whether the universe was created or whether it will vanish. For in such a case, I would have to ask the cause of the creation or the whereabouts of the creator falling into an infinite regression. And also I would have to ask what happen with the M after it has vanished; for it is not possible for me to conceive that after such event only empty space and time remained. I believe that it is absurd to think that space exists only as a container of M without thinking that space itself is made up of some kind of M. And also for time, time would have no meaning if matter were not constantly changing. By this I champion Aristotle's view wherein is stated that motion precedes time \cite{aristoteles}; the cause that we believe in a flow of time is motion or change of matter. The change of matter relative to matter itself makes us feel that something that we have called \emph{time} flows. We perceive different events (visual, auditive, etc.) and thanks to our material memory we feel a flow of time because we compare a current event with a previous one. Similarly, our notion of Newtonian space arises in relation to material objects. For instance, if I remove say an apple from a table, my brain immediately tells me that moments ago there was something there filling certain volume and occupying some place or position, because in relation to the other objects in the room, which still remain in the same position, there is something missing. Hence, I think that space remain there but the object not. Now if we further imagine that we remove all objects in the room, then even the room, the earth, the stars, galaxies and so on, all things in the universe, we are left only with space and time (or nothing if you wish), but how can space (or nothingness) exists if is not made up of something? How can time flow if nothing changes?
  
 Some others may argue that motion is referred to space and time. Yes, but how do we measure space and time? The way we measure time is by motion or change of matter, and the way we measure space is in relation to material objects. An electronic clock (or any other kind) is an instrument that is continually changing; each second is a completion of a process that is taking place inside the ``machinery" whose parts are made up of matter in motion, if there were no changes in the clock a process would not be completed and a second in the clock would not be displayed. Does this imply that time does not flow? Certainly, time still flows because time is the intrinsic motion and change of the universe, motion and change can never stop. And a ruler is a material object we use to compare and delimit a particular length, without matter, space would be meaningless too, for there would be nothing to relate the motion. For such reason space itself should be a material continuum even if there were no ponderable objects to refer.
  
  \item Since nothing can be created out of nothingness, it cannot be empty spaces where there exists nothing, which implies that the universe and space are made up of continuous M. Hence, there is no room for material discontinuity, total emptiness is inconceivable to me. A volume can be deprived of ponderable matter but not from imponderable matter. Thus, from here we must also conclude that imponderable matter must constitute a continuous medium in conjunction with ponderable matter. And also for the description of physical phenomena imponderable matter might be seen as an absolute physical reference frame because matter evolves relative to matter; physicists know that what I have stated may imply the abandonment of the philosophy of the general relativity \cite{jacobson}, though this does not necessarily force us to give up the covariance of the physical laws (please see \cite{perez1,christov7,christov5,christov1,christov6}). What I have said simply implies that we are living immersed in a \emph{dynamical material space}.
  
To even further support my notion about space let us consider the following paradox of place as put forward by Aristotle:
\begin{description}
  \item[] 
 If everything that exists has a place, place too will have a place, and so on \emph{ad infinitum}.
\end{description}
The premise in which the Aristotle's statement is founded is the assumption that space exits, but nevertheless, it is implicit that space is not made up of anything, which is contradictory. For this reason one arrives at the fallacy that everything that exists, including place, must have or occupy a place. Now recall that for ancient thinkers, \emph{matter} meant: \emph{space occupied}. Hence the paradox is resolved when one acknowledges that space is made up of matter, and therefore space cannot have or occupy space \emph{ad infinitum}. 
  
  \item I also hold the position that ideal objects are part of the universe. Of this kind are mathematics and any other hallucination, dream or idea created by my own being because reasoning is the product of the dynamics of the universe. And although what I think dissipates energy, this does not entail that what I am thinking 'exists' or 'is' in the real (or measurable) universe. Only the laws of logic as well as the laws of experience \cite{mittelstaedt} will dictate whether the interpretations that I have constructed to describe the universe are univocal to it. And again, for this reason, if space is some physical entity, and therefore exists it must be made up of something, otherwise it is an element created by my imagination with no physical constitution. Of that nature is topology or Euclidean geometry which epitomizes the Newtonian background of space.
  
  To deeply understand this consider the following reflexion which will expose very vividly the nuances between the mathematical objects and the physical ones:
\begin{description}
  \item
\emph{If a physical object which is not made up of something exists one has an object that is physically nothing and therefore, in spite of this, one is able to put an infinitely number of these objects in the same place. At the end of the task, since these objects are nothing, we end up with the same object, that is, with nothing. So we rise the question: what is the need of having a physical object that exists but nevertheless is physically nothing?}
\end{description}
From here we evidence that mathematical objects can exist in our imagination, in a certain sense, emerge out of nothing and at a given moment become nothing, vanish; but can a physical object behave in the same way? can physical space exist and be made up of nothing? If not, can space disappear? I do not believe so. 
 
    \subsection*{There is neither temporal beginning nor temporal end of the universe}
  
  \item I think that the universe is, exists, has always been and will exist indefinitely and infinitely, the universe will never become into nothingness. I retain the opinion that there was no moment of creation and it will not be an end. For thinking of the occurrence of these events simply implies a change from a particular state to a distinct one, a simple transition. In a similar way to the points of a circumference in which any arbitrarily chosen point can be the beginning of the circumference, in the same way occurs with the universe, the beginning or end is mere convention to delimit two major events. Even more, each moment can be the beginning or end of a series of events; this is what is called \emph{evolution}, where \emph{causality} is implicit. Hence, the problem to be resolved is the \emph{principle of causality}, which demands an initial cause (initial conditions in mathematical terms) and, at the same time, lead us to an infinite regression, for we must ask the cause of the Big Bang (if we believe in this model) and so on and so forth. Therefore, we should accept that inside the notion of time the dynamics of the universe is involved, time is the word we use to denote the changes that the universe suffers and is also the word use to determine the movements of bodies, its other name is not precisely \emph{duration} \cite{newton1} but rather \emph{mutation}, \emph{material change}. Both time and the principle of causality constitute our second great problem to be studied. Since material change is the source of the notion of time the big question to be answered is not merely ``\emph{what is time?}" but rather, ``\emph{why things change?}". I think that once one has recognized that the universe 'is' or 'exists', one must ask: why does the universe evolve on its own? what motor propels the universe? is there really an initial cause that started the motion of the universe?
  
\subsection*{The universe must be infinite}
  
    \item And when in paragraph 2 we talked about ``external" or ``internal", these ideas must be referred to something that separates or delimits two things or sections where there is something contained. To speak of ``outside" and ``inside" there must be a boundary or a criterion to demarcate the limits. If the boundary existed such adjectives would be conventional and would depend on which side of the boundaries we were. If the boundary did not exist there would not be the possibility of distinguishing one from the other. But the boundary, we must acknowledge, must be material, since we have rejected the existence of nothingness. If the boundary existed we would be able to distinguish one thing from another and one event from another; this is what we call \emph{discretization}. However, this does not imply that the universe is made up of discontinuous matter.
    
    \item Apparently, the intellect has no boundaries, one can imagine whatever one pleases \footnote{Here I appeal to the potential or conceptual infinite \cite{aristoteles,rucker,moore3}}. Thus, ideally or mathematically the existence of a boundless and finite universe is possible, but in physical terms this is inconceivable and contradictory. For a physical and finite universe must have physical boundaries and therefore it must contain finite quantities of matter and energy. If the universe has boundaries, they must be of M to delimit it and isolate it. But, isolate it from what? From other type(s) of universe(s) that is (are) next to or surrounding it? or isolating it from an exterior space? If such boundaries exist they must interact with the neighbor universe(s) or the exterior space(s), hence, these universes must be part of a superuniverse of infinite extension, therefore there cannot be physical limits. It follows that \emph{the universe must be infinite in extension \footnote{Here I make reference to the actual infinite.} and therefore the global quantities of matter and energy must be also infinite}.
        
  \item From the previous reflexions it follows that there is only one universe and, hence, the ideas of  physical multiverses appear to be absurd and contradictory, although, from the mathematical point of view any assertion of this kind can be useful to get rid of other mathematical complications. Thinking of the existence of physical universes that do not interact with our universe makes no sense to me, for we would have to admit that what divides those universes is not physical (or of whatever nature) that allows us the interaction \cite{smolin,stoeger,tegmark2}. If these universes exist in one way or another they must interact with ours and among them. From the perspective exposed here denying their interaction is incoherent. If their interaction is tiny or huge that is quite different. If the types of matter that pervade those volumes were different than the one pervading ours we would expect a different dynamics and possibly different values for the physical constants.
  
  \item Einstein speculated on these issues \cite{einstein11,einstein12}; it was no possible for him to conceive that the radiation emitted from celestial bodies get lost forever in the immensity of space. Based on the definition of density $\rho=m/V$, where $m$ is the averaged mass of the universe and $V$ its volume, he proposed the following criteria to determine the finiteness or infiniteness of the universe.
   \begin{enumerate}
  \item If $\rho=a$, where $a$ is a positive number, then the universe is finite.
  \item If $\rho=0$ the universe is infinite.
\end{enumerate}
From the first option it is easy to deduce that both mass and volume are finite. Here the density that Einstein considered is the density of ponderable matter, that one due to electromagnetic radiation and the density by the gravitational binding energy. In such case the contained quantity of matter and energy is finite and it is evident that in a universe of this kind (close and elliptic curvature) any physical quantity is conserved and therefore the laws of conservation (charge, mass, energy, momentum, etc.) turns out to be a tautology.

The second case can only take place if the volume is infinite since it is evident that ponderable mass is not zero. But this line of argumentation implies the assumption of an infinitely empty space (vacuum) with no material constitution at all, which, we have stated, is absurd.

Thus, based on our materialism, the interesting thing comes now, since in items 12 and 13 we have concluded that a finite universe must have physical boundaries, but this is also contradictory then we must reject option (a) and we have no other option but to conclude that the universe is infinite. But if the universe is infinite in extension we have also said that it must be infinite in M, therefore we must also reject option (b) and the expression becomes $\rho=\infty/\infty$. This result is not to be discussed within the realm of ordinary functional analysis but within the context of either non-standard analysis or the surreal context which may provide us some insight into the nature of infinite quantities (see for instance \cite{vladimir,woodin,robinson,goldblatt}).

At this point I  must confess the impossibility for discussing something about the  conservation laws (symmetries) in a universe with infinite extension, mass and energy. Even more, one can claim that a body of infinite dimensions is inconceivable for such an object should then posses infinite matter and energy, which at first sight is also absurd. In other words, a theory like this may not be testable and, consequently, if a theory is not falsifiable \cite{popper2}, sounds like metaphysics. In this favor Stoeger \cite{stoeger} argues that ``something which is unboundedly large, and therefore not specifiable or determinate in quantity or extent, is not materially or physically realizable". Hence, in this way arises our third great problem to be solved: \emph{the infiniteness of the universe and its implications in the conservation laws}. 

\item In item 7 I mentioned that everything is material, then from items 11-14 I argued that the universe should be infinite in extension, and that movement, which causes the notion of time, has always existed and it will never end. In other words, once the universe is realized, the universe is infinitely gifted, in mass, extension, energy and perpetual continuous change (time). Are these preliminary conclusions paradoxical? I do not see such. For I cannot acknowledge that the motion of the universe began out of no motion, out of no change and that matter was created out of nothingness. For if I would, and thus following Kant, it would also be legitimate to acknowledge the antithesis that matter exists but it is made up of nothing, that matter is a void which seems to me quite contradictory. And similarly I would have to admit that matter spontaneously came into being at total rest and, by unknown causes, spontaneously started to move. And therefore in the reverse order the universe will become nothingness.

\end{enumerate}

\section{Identity and regularity: the laws of physics}
\begin{enumerate}
  \item Up to now I have never experienced identical or regular things. It has been said that identity and regularity can only be assumed but never grasp by experience. The clear example of identical objects are the mathematical ones, e.g., two triangles; of the regular, the physical laws. Since the objects of mathematics can be identical we think that also its structures, postulates, axioms, symmetries, etc. appear in the physical world. And we search such qualities in nature even though we have never experienced them. For it is said that a sphere has a rotational symmetry, but has anyone perceived a physical object that be a perfect sphere? And it is also thought that the laws of nature are regular because, for instance, we observe motion patterns of the planets, but who can assert with total certainty that the earth follows such or such trajectory? Heisenberg's uncertainty principle forbids us the exact determination of such motion, but in spite of that we assume it. And although we know that the objects of the universe are in constant mutation and that the measuring instruments we used to interact with the universe are constantly changing too, we suppose, in spite of this, that the laws that rule the universe are regular and immutable. Even more, we search for regular patterns, symmetries and identical forms where we know that there are not, because we have never experienced them, but we only assume them. This perplexity is always present in the mind of the experimental physicist for he knows that the error bars will never become zero. For the theoretician, however, physics is ``the art of approximations". As implicit in Heisenberg's principle, nature has lost its causal relation, the principle of causality has no meaning anymore for there is no way to assert the exact position or momentum of a particle. So there is no sense in speaking about a reality that can never be grasped by experience (measurements).
  
  \item Our theories are, thus, only assumed realities, models that gradually enhance their structures by continuous interaction (experience) with the universe. From the previous lines we can figure out our four great problem: are the laws of physics truly regular? Some physicists argue that we only observe effective (phenomenological) laws and not genuine laws of nature. But are the laws of physics predetermined if the universe is in constant mutation? The tendency of my thought inclines in the same way as the universe behaves, that is, the laws of physics are constantly changing, though in our short existence, in comparison with the time scale of the universe, they do it very slowly. Because of space is material for us and is the medium for the propagation of electromagnetic waves, it is natural to think that the space of some millions ago possessed different physical properties than those we observe today. For, it is well known that, the absolute speed of a wave in a medium with different physical properties (e.g. non constant refractive index) will show, accordingly, different values. This argument, therefore, applies to both physical constants and laws; hence, in a universe with other physical conditions other laws (symmetries) must apply.
  
  \item If the universe has no physical laws the work of the professional physicist will be in vain. His attempt to formulate the behavior of the universe in laws of nature will become a stubborn chimera. And as we have stated in item 4 above, we are uncertain whether knowledge is truly possible and, if so, whether total knowledge is truly realizable. Some argue that the function of a law is not to prescribe the behavior of a system but to describe it. In any case we are subjected to the laws not merely of nature but of logic and coherence as well, to the laws of reasoning \cite{mittelstaedt}. But if reasoning and logic are the consequence of the evolution of the universe, one is then forced to admit that the laws of logic and reasoning are laws of nature.
 
\end{enumerate}

\section{Discussion}
\label{sec}

 From what has been propounded in the previous sections, what things can be useful to address the problems I listed in section \ref{trou}?. First, that assuming the whole universe made up of one single entity there are no arguments to state problems 3-8. Hence we do not have to be surprised of dark matter or dark energy. The assumption that the universe is made of two different entities, matter and fields occupying empty space turns out to be redundant. Fields and particles (of the standard model) can be seen simple as states of that subtle matter (see the similarities with condensed matter \cite{volovik}). In these respects C. Christov \cite{christov7,christov5,christov1,christov6} 
has introduced the notion of phase patterns from the theory of solitons to provide an explanation of how the quantization of matter arises out of the material continuum. In other words, a particle can be seen as a soliton (or quasi-particle). In this way problem three can be addressed. Following this approach he has also achieved a unified theory of gravity, electromagnetism and wave quantum mechanics; as a consequence the wave-particle duality fades away. The \emph{special theory of absolutism} (STA), as the author called it, is conceptually and mathematically much simpler than the most popular unified theories, i.e., M-theory or loop quantum gravity (LQG). And by this approach one can address problems 1-7. Problem 8 has already been discussed and within the latter theory there is no need for several universes since they become superfluous; moreover STA also explains in what sense one can adopt another spatial dimensions. Another modern proposal to address the second problem can be found elsewhere \cite{griffiths,hohenberg}.

Also from our above discussion, one can easily deduce that the universe is not symmetric before time reversal since it is continuously changing by itself, the cause of the changes yet unknown. Experience and common sense however suggest us that two identical processes cannot repeat twice in exactly the same form, therefore, no process can be reversible. This is stated thermodynamically by saying that the entropy of the system increases. On the other hand, reversibility implies causality, and in this principle is tacit that the same initial conditions give as result the same effects. A reversible process is an idealization in which one assumes that the principle of causality always holds.  But in a dynamical universe it is impossible to guarantee the same initial conditions, we can only assume them; it follows that the same effects are not expected. Hence the classical arrow of time appears as consequence of the incomplete mathematical formulation of the physical laws. For instance, Fritz Rohrlich has shown that when one considers the self-field (gravitational or electromagnetic) \footnote{Most of the times this field corresponds to a test charge so minute that is commonly neglected.} of a particle the classical equations of mechanics become asymmetric before time reversal \cite{gold,rohrlich1,rohrlich3,medina}. And therefore the problem of the classical arrow of time becomes fictitious.

As to problems 11 and 12, it is instructive to bring the words of John Bell into this context when he alluded to the notion of measurement \cite{bell3}. He declared: 

\begin{description}
  \item \emph{Why should the scope of physics be restricted to the artificial contrivances we are forced to resort to in our efforts to probe the world? Why should a fundamental theory have to take its meaning from a notion of ``measurement" external to the theory itself? Should not the meaning of ``measurement" emerge from the theory, rather than the other way around?}
\end{description}
Since the very moment that physics was regarded as an experimental science, a physical theory is doomed to measurements. Thus, any propounded theory must have constants or parameters to be determined by experiment. This is the case of $G$ in Newtonian gravitation, of $\epsilon_0$ and $\mu_0$ in Maxwell electrodynamics, this is also the case of $c$ in special relativity (SR), of $h$ in quantum mechanics, of the $n$ parameters in the standard model of particles and of the $m$ parameters in the standard model of cosmology. But why the constants have that value, it seems to me, that this question is not well formulated. For instance the speed of light in vacuum was conventionally defined by the Bureau Intertanational des Poids et Mesures (BIPM) as $V_r=$ 299 972 458 m/s. But this value was taken as a convention \cite{giacomo0}, this does not imply that the actual (or measured) speed of light possesses that exact value but the actual value is around $V_r$ with a speed uncertainty of $a<1$ m/s. By convention, again, $V_r$ was identified with the parameter $c$ of SR \cite{ellis,mendelson}. However, the theoretical value, say $V_r=1/\sqrt{\mu_0\epsilon_0}$ that stems from Maxwell's electrodynamics and represents the speed of the light waves in vacuum has been given the experimental value 299 972 458 m/s that corresponds to a measurement in the radiation zone. But Budko has shown that Maxwell's solutions in vacuum in the near and intermediate zone allow values different from $V_r$, say, $V_n$ and $V_i$, respectively \cite{mugnai,budko}. By the principle of relativity Maxwell equations must be invariant in any other inertial frame and hence Maxwell's solutions for the near and intermediate zone must be valid as well, no matter if $V_n>V_r$ and/or $V_i>V_r$. Now we ask: are we violating the second postulate of relativity? Is the parameter $c$ really equal to $V_r$? Why not $c$ was taken to be equal to $V_n$ or $V_i$? Recall that for SR to make physical sense, the parameter $c$ must be higher than the speed of the inertial frame $v$, so that the Lorentz transformations do not render complex numbers. In this sense the selection $c=V_r>v$ is partially justified, but we could have conventionally defined $c=$299 999 999 m/s and the physics would not be affected at all since the theory by itself only demands a constant with units of speed with any value but different than $v$. But why did SR borrow the value from another theory (electrodynamics)? Why is not SR capable of determining the value of its own constants? The theory then, with no relation to a measurement, cannot determine the value of $c$ by itself. These arguments also apply for any other theory (see for instance \cite{narlikar}). If a theory were able to determine the values of its constants and parameters, the theory would likely become independent of experience (measurement) \cite{tegmark5,tegmark1}.

Finally, if nature imposes barriers to be knowledgeable, which are implied in the uncertainty principle and the laws of statistics, that is, the error bars will never become zero, and if we add the idea that total knowledge of the universe is not possible, we might conclude that no final theory will ever be achieved. In this way the last question may be answered.

\section{Concluding remarks}
\label{conclu}

In summary, we have discussed the constitution of the universe which answer the question: What is space made up? The consequence of realizing space as something material forces us to abandon the notion of nothingness and total emptiness as well as the necessity of the multiverse as physical realities. The fundamental questions that are left to answer are: What are matter and energy? What are their fundamental properties, their types and how they interact, once we have unified the concepts of fields and particles as part of one single entity? The great second question is: What is time? Since here is where we found the dynamics and evolution of the universe as well as the principle of sufficient reason. LQG has realized the importance of studying the nature of time \cite{markopoulou}. The third great problem that we have isolated is the infiniteness of the universe, from which we have to investigate the implications in the symmetries of nature \cite{volovik}. Finally, we put the problem of the regularity of the laws of physics on the table. In our theories we have simply assumed that in the past and in the future the laws of physics were and will be the same ones that we found today, an assertion that, in fact, is just matter of faith but not an experimental fact, although for practical purposes, the fact that they were regular or not may turn out to be irrelevant. Instead we may only talk about interim laws of physics, interim symmetries, laws that are only valid within certain periods of time, after some time they become obsolete because the universe is no longer the same. 

\begin{acknowledgements}
The author is grateful to Georgina Carrillo. A CINVESTAV grant is acknowledged.
\end{acknowledgements}

\end{document}